\documentclass[pra,superscriptaddress,showpacs,preprint,aps]{revtex4-1}

\usepackage{epsfig}             
\usepackage{epstopdf}           
\usepackage{hyperref}           
\usepackage{color}
\usepackage{colortbl}
\usepackage{graphicx}
\usepackage{amssymb,amsmath}
\usepackage{subfigure}
\usepackage{caption}
\usepackage{enumerate}
\usepackage{gensymb}
\begin{document}

\title[WB]{Wannier-Bloch approach to localization in high harmonics generation in solids}

\author{Edyta N. Osika}
\email[]{edyta.osika@fis.agh.edu.pl}
\affiliation{AGH University of Science and Technology, Faculty of Physics and Applied Computer Science, al. Mickiewicza 30, 30-059 Krak\'ow, Poland}
\affiliation{ICFO - Institut de Ci\`encies Fot\`oniques, The Barcelona Institute of Science and Technology, 08860 Castelldefels (Barcelona), Spain}

\author{Alexis Chac\'on}
\email[]{alexis.chacon@icfo.es}
\affiliation{ICFO - Institut de Ci\`encies Fot\`oniques, The Barcelona Institute of Science and Technology, 08860 Castelldefels (Barcelona), Spain}

\author{Lisa Ortmann}
\affiliation{Max-Planck Institut f\"ur Physik komplexer Systeme, N\"othnitzer-Str. 38, D-01187 Dresden, Germany}

\author{Noslen Su\'arez}
\affiliation{ICFO - Institut de Ci\`encies Fot\`oniques, The Barcelona Institute of Science and Technology, 08860 Castelldefels (Barcelona), Spain}

\author{Jose Antonio P\'erez-Hern\'andez}
\affiliation{Centro de L\'aseres Pulsados (CLPU), Parque Cient\'{\i}fico, E-37185 Villamayor, Salamanca, Spain}

\author{Bart\l{}omiej Szafran}
\affiliation{AGH University of Science and Technology, Faculty of Physics and Applied Computer Science, al. Mickiewicza 30, 30-059 Krak\'ow, Poland}

\author{Marcelo F. Ciappina}
\affiliation{Max-Planck-Institut f\"ur Quantenoptik, Hans-Kopfermann-Str. 1, 85748 Garching, Germany}
\affiliation{Institute of Physics of the ASCR, ELI-Beamlines, Na Slovance 2, 182 21 Prague, Czech Republic}

\author{Fernando Sols}
\affiliation{ICFO - Institut de Ci\`encies Fot\`oniques, The Barcelona Institute of Science and Technology, 08860 Castelldefels (Barcelona), Spain}
\affiliation{Departamento de F\'isica de Materiales, Universidad Complutense 
de Madrid, E-28040 Madrid, Spain}

\author{Alexandra S. Landsman}
\affiliation{Max-Planck Institut f\"ur Physik komplexer Systeme, N\"othnitzer-Str. 38, D-01187 Dresden, Germany}

\author{Maciej Lewenstein}
\affiliation{ICFO - Institut de Ci\`encies Fot\`oniques, The Barcelona Institute of Science and Technology, 08860 Castelldefels (Barcelona), Spain}
\affiliation{ICREA - Instituci\'{o} Catalana de Recerca i Estudis Avan\c{c}ats, Lluis Companys 23, 08010 Barcelona, Spain}

\date{\today}

\begin{abstract} 
Emission of high-order harmonics from solids provides a new avenue in attosecond science. On one hand, it allows to
investigate fundamental processes of the non-linear response of electrons driven by a strong laser pulse in a periodic crystal lattice. On the other hand, it opens new paths toward efficient attosecond pulse generation, novel imaging of electronic wave functions, and enhancement of high-order harmonic generation (HHG) intensity. A key feature of HHG in a solid (as compared to
the well-understood phenomena of HHG in an atomic gas) is the
delocalization of the process, whereby an electron ionized from one site
in the periodic lattice may recombine with any other.  Here, we develop an analytic model, based on the localized
Wannier wave functions in the valence band and delocalized Bloch functions in the conduction band.  This Wannier-Bloch approach assesses the contributions of individual lattice sites to the HHG process, and hence addresses precisely the question of localization of harmonic
emission in solids.  We apply this model to investigate HHG in a ZnO
crystal for two different orientations, corresponding to wider and narrower valence and conduction bands, respectively.  Interestingly, for narrower bands, the HHG process shows significant localization, similar to harmonic generation in atoms.  For all cases, the delocalized contributions to HHG emission are highest near the band-gap energy.
Our results pave the way to controlling localized contributions to HHG in a solid crystal, with hard to overestimate implications for the emerging area of atto-nanoscience. \end{abstract}

\maketitle

\section{Introduction}
Recently, the techniques of attosecond science, traditionally applied to atoms and molecules in the gas phase~\cite{Krausz2009}, have been extended to solid state ~\cite{ leone2014, schultz2014, Schiffrin, Ghimire, luu2015, huber2015, Vampa2015, Reis2016,NepplPRL, Cavalieri,Neppl}.  A crucial difference between solid and gas targets is the localization of the initial state electron wave function, which is spatially confined in isolated atoms and molecules, but can be delocalized in a solid.  The effect of  wave-function localization on key aspects of light-solid interaction remains intensely debated.
Hence some attosecond experiments~\cite{Cavalieri,Neppl} on photoemission from metal surfaces suggest that the localization of the core-band electrons results in relatively large ionization delays, attributed to transport~\cite{Zhang}, compared to photoemission from delocalized conduction-band states. Other experiments probing photoemission from the same initial state at different photon energies found that larger ionization delays came from resonant excitation into bulk excited states, rather than from the initial localization of the wave function~\cite{Optica,ScienceMurnane}.\\
In this work, we investigate electron localization and the underlying microscopic nonlinear response by focusing on the process of high harmonic generation (HHG) in a crystal solid. HHG, a cornerstone of attosecond science, has traditionally relied on gas phase atomic targets, and has only recently been demonstrated experimentally in condensed phase~\cite{Ghimire,luu2015,huber2015,Vampa2015,Schubert2014,Reis2016}.
A key feature of HHG in atoms is the recombination of the ionized electron with its parent ion, making it a highly localized process.  This localization both dramatically limits HHG efficiency and leads to an exponential decline of HHG yield with increasing ellipticity of laser light  (ellipticity creates a drift, which exponentially suppresses the return of ionized electrons to the parent ion~\cite{Burnett}).  

In contrast, the HHG process in a solid can be delocalized, since an electron ionized from one site of the crystal lattice may recombine with any other. However, little is understood about the specifics of this process.
For instance, Ghimire {\em et al.}~\cite{Ghimire} found a much weaker dependence of high harmonic yield on ellipticity in solid ZnO than would be expected for a gaseous medium, suggesting a highly delocalized process.  A significantly stronger ellipticity dependence (although still weaker than in atoms) in the same target was subsequently found in a theoretical work~\cite{Brabec}, which shows a 2-3 orders of magnitude drop in HHG yield for ellipticity of 0.5 (compared to only a factor of 5 drop measured in Ghimire~\cite{Ghimire}).  At the same time, a recent experiment on solid Argon found the same dependence on ellipticity as in gas-phase Argon, suggesting the electron recombines with the same lattice site that it was emitted from~\cite{Reis2016}.  The extent of spatial localization, measured experimentally by ellipticity dependence, is believed to be important for attosecond pulse generation and imaging of the electronic wave functions in the solid state~\cite{Reis2016}.

Here, we investigate the spatial dependence of the HHG process in ZnO by introducing an analytic model which uses localized Wannier wave functions in the valence band and delocalized Bloch functions in the conduction band.  
Prior seminal work~\cite{Brabec, Vampa_bands,GhimirePRA2012} used delocalized Bloch functions both in the valence and conductions bands, and hence was not able to extract spatial information.
In addition to accurately calculating the total HHG yield, this Wannier-Bloch approach allows us to separate the contributions of individual lattice sites to each harmonic, and hence determine the degree of localization of the HHG process as a function of experimental parameters.  We find that this localization varies significantly both with the harmonic order and with the orientation of a crystal.  
Our results point to a possibility of controlling the spatial localization of the HHG process (by varying, for instance, laser ellipticity or crystal orientation), with hard to overestimate implications for attosecond pulse generation, HHG efficiency,  imaging of attosecond electron dynamics in condensed matter, and for the emerging area of atto-nanoscience as a whole~\cite{reviewROP}.

\section{Wannier-Bloch description of the high-order harmonic generation in solids}
Analogous to the three-step model in atoms \cite{Corkum1993}, the HHG in a crystal solid via interband transitions can be described as follows \cite{Brabec}:  ({\em i}) electron (hole) tunneling excitation from the valence band to the conduction one, ({\em ii}) electron (hole) acceleration in the conduction (valence) band, and ({\em iii}) electron-hole recombination, resulting in an emission of a high harmonic that is a multiple of the frequency of the driving laser.  

In most recent experiments the laser field strength, ${\mathcal E}_0$, across the lattice
constant $a$ is comparable to the band-gap energy $E_g$ of a typical semiconductor (${\mathcal E}_0a\simeq E_g \simeq$ few eV). As a consequence, the field can not be considered as a small perturbation \cite{Chin2001}. 
In our model we therefore assume that this condition is satisfied, but the laser field amplitude is below the damage threshold.
In addition, the photon energy of the laser field should be much smaller than the typical bandgap energy. This means we restrict our studies to the photon energies in the MIR domain ($\hbar\omega_0 \leq 0.5 $~eV), which implies that the central-laser wavelength $\lambda_0$ is much larger than the typical lattice constant, $a$. Thereby, the dipole approximation is valid for our description of laser-solid interaction.
Since the laser field is linearly polarized in the $x$ direction, we adopt here a one-dimensional description.
The Hamiltonian of a single electron in a crystal under the action of a laser field is given by
\begin{equation}
H(t)=H_{0}+U_{\rm int}(x,t), \label{Eq:FullHam}
\end{equation}
where
\begin{equation}
H_{0}=-\frac{1}{2}\frac{\partial^2}{\partial x^2}+U(x)
\end{equation}
is the laser-free Hamiltonian, with $U(x)$ the lattice periodic potential. In Eq.~(\ref{Eq:FullHam}), $U_{\rm int}(x,t)=-q_exE(t)$ is the oscillating potential due to the laser, written in the length gauge. Here we use atomic units $\hbar=|q_e|=m_e=1$, where $q_e$ and $m_e$ are the electron charge and electron mass, respectively. The laser pulse has the form $E(t)={\mathcal E}_0\sin^2(\omega_0 t/2N)\sin(\omega_0 t+\varphi_{\rm CEP})$ where ${\mathcal E}_0$ is the electric field peak amplitude, $\omega_0$ the carrier frequency and $\varphi_{\rm CEP}$ the carrier-envelope phase (CEP) of the laser field, while $N$ is the number of laser-period cycles.\\
Unlike the prior work ~\cite{Brabec, Vampa_bands}, we describe the system within a mixed representation: Wannier states in the valence band and Bloch states in the conduction band. In contrast to the Bloch functions, the Wannier functions are spatially localized ``elements" of an $L^2$ space.  
In terms of localized wave functions, they provide thereby an analogous insight into HHG mechanism as the usual approach used in atomic and molecular systems. Furthermore the Wannier functions form a complete orthogonal set in the valence band, but are not eigenfunctions of the Hamiltonian $H_0$. In our problem the initial state corresponds to a completely filled valence band, i.e.~a completely filled Fermi sea. This means that initially all the Bloch states are occupied or, equivalently, that all Wannier states are occupied. We have thus to solve the time dependent Schr\"odinger equation starting with each Wannier state, and sum up the results at the end. We introduce an ansatz for the complete time-dependent states of a single electron in a lattice as a superposition of Wannier states $|w_{v,j}\rangle$ from the valence band and Bloch states $|\phi_{c,k}\rangle$ from the conduction band
\begin{equation}\label{eq:wavefunction}
|\Psi(t)\rangle=\sum_{j}|w_{v,j}\rangle a_{j}(t)+\int_{BZ}a_{c}(k,t)\,|\phi_{c,k}\rangle dk,
\end{equation}
with the initial condition $a_j(0)=\delta_{j,j_0}$, i.e.~the electron starts the dynamics at the site $j_0$.
Here $j$ runs over all atomic sites in the crystal. The Bloch functions of an $m$-th band ($m=v$ for valence band and $m=c$ for conduction band) have a form
\begin{equation}\label{eq:Bloch}
\phi_{m,k}(x)=u_{m,k}(x)e^{ikx},
\end{equation}
where $u_{m,k}$ is a periodic function with the same periodicity as the crystal. The wave functions in Eq.~(\ref{eq:Bloch}) can be equivalently represented by a set of Wannier functions
\begin{equation}
w_{m,j}(x)=\int_{BZ}\phi_{m,k}(x-x_{j})\tilde{w}_{m}(k)dk,
\end{equation}
where $\tilde{w}_{m}(k)$ is a product of a normalisation constant and a phase depending on electron momentum $k$. It has been shown in \cite{Kohn} that for a 1D
lattice the $\tilde{w}_{m}$ are independent of $k$ provided the Wannier functions are real and symmetric under appropriate reflection and fall off exponentially with distance. 
So, to calculate the emitted harmonics firstly we compute the time-dependent dipole moment
\begin{eqnarray}
d(t)&=&-\langle\Psi(t)|x|\Psi(t)\rangle\nonumber\\ 
&\approx&-\int dx\int_{BZ}dk\sum_{j} x w^*_{v,j}(x) a^*_{j}(t)\phi_{c,k}(x) a_{c}(k,t)+c.c.\nonumber\\ 
&=&\int_{BZ}dk\sum_{j} a^*_{j}(t) d_{jc}(k) a_{c}(k,t)+c.c., \label{Eq:DipoleRad}
\end{eqnarray}
where $d_{jc}(k)$ is a dipole transition matrix between Wannier $w_{v,j}(k)$ and Bloch $\phi_{c,k}$ states. The physical meaning of this equation can be summarized as follows: at the observed time, $t$, the electron previously promoted to the conduction band recombines with the valence band via $d_{jc}(k)$ and emits a photon with an amplitude which depends on the amplitudes $a_j(t)$ and $a_c(k,t)$. 
Secondly and similar to Vampa {\em et al.}~\cite{Brabec}, the harmonic emission is obtained by modulus squared of the Fourier transform of Eq.~(\ref{Eq:DipoleRad})
\begin{eqnarray}
I_{\rm HHG}(\omega)&=&\omega^2|\tilde{d}(\omega)|^2 \nonumber\\
\tilde{d}(\omega)&=& \frac{1}{\sqrt{2\pi}}\int_{-\infty}^{\infty}dt\,e^{i\omega t}\,{d}(t)\,.  \label{Eq:DipoleSpectrum}
\end{eqnarray}
According to Vampa {\em et al.}~\cite{Brabec}, at long-laser wavelengths, i.e. between 1.0 and 5.0~$\mu m$, the main contribution to the harmonic spectrum is from interband transitions.  Consequently, we neglect the intraband contribution terms $\langle w_{j,k}|x|w_{j,k'}\rangle$ and $\langle \phi_{c,k}|x|\phi_{c,k'}\rangle$ in Eq.~(\ref{Eq:DipoleRad}).  Our main task then consists of computing the dipole transition $d_{jc}(k)$ and the transition amplitudes $a_c(k,t)$ and $a_j(t)$. 
The dipole moment $d_{jc}$ can be further expressed as a product of two terms: one dependent and one independent of $j$. First, following \cite{Brabec} we approximate the matrix elements as follows
\begin{eqnarray}
\langle\phi_{c,k}|x|\phi_{c,k'}\rangle&=&i\nabla_{k}\delta(k-k'),\\
\langle\phi_{c,k}|x|\phi_{v,k'}\rangle&=&-d_{cv}(k)\delta(k-k')
\end{eqnarray}
with $d_{cv}(k)=-\langle\phi_{c,k}|x|\phi_{v,k}\rangle$. The transition dipole moment from conduction to valence band is then expanded
\begin{eqnarray}
d_{jc}(k)&=&-\int dx w^*_{v,j}(x) x \phi_{c,k}(x)\nonumber \\
&=&-\int dx \int_{BZ}dk' \phi^*_{v,k'}(x-x_{j})\tilde{w}^*_{v}\,(x-x_j) \phi_{c,k}(x-x_j)e^{ikx_j} \nonumber \\
&=&d_{vc}(k)\tilde{w}^*_{v}e^{ikx_j}.
\end{eqnarray}
The replacement of $x$ by $(x-x_j)$ in the above formula is justified by the fact that $\langle w_{v,j}|x-x_j|\phi_{c,k}\rangle = \langle w_{v,j}|x|\phi_{c,k}\rangle$.
In addition, to obtain the coefficients $a_j(t)$ and $a_c(k,t)$ we employ the time-dependent Schr\"odinger equation, Eq.~(\ref{Eq:FullHam}), with the wave functions defined in Eq.~(\ref{eq:wavefunction})
\begin{equation}\label{eq:TDSE}
i\frac{\partial}{\partial t}|\Psi(t)\rangle=H(t)|\Psi(t)\rangle.
\end{equation}
We use the tight-binding approximation for the description of the band structure and assume the dispersion relations for the valence/conduction bands
\begin{eqnarray}
E_v(k)&=&-2I_v\cos(k a) \nonumber \\
E_c(k)&=&E_c'-2I_c\cos(k a)
\end{eqnarray}
where $I_v$ and $I_c$ are hopping parameters in the valence and conduction bands respectively, $a$ is a lattice constant, $E_c'=E_g+2I_c-2I_v$ and $E_g$ is the bandgap energy of the solid material. 
The matrix elements for the unperturbed Hamiltonian in both the valence and conduction bands read
\begin{eqnarray}
\langle w_{vj}|H_0|w_{vj'}\rangle &=& -I_v\delta_{|j-j'|,1} \nonumber \\
\langle \phi_{c,k}|H_0|\phi_{c,k'}\rangle&=&E_c(k)\delta(k-k'),
\end{eqnarray}
respectively.
Thereby, with the previous definitions and after introducing the wave functions Eq.~(\ref{eq:wavefunction}) into Eq.~(\ref{eq:TDSE}), we end up with a system of coupled differential equations for $a_j(t)$ and $a_c(k,t)$
\begin{eqnarray}\label{eq:ajdot}
\dot{a}_j(t)&=&iI_v {a}_{j-1}(t)+iI_v {a}_{j+1}(t)-ix_jE(t)a_{j}(t)+iE(t)\int_{BZ}dk d_{jc}(k)a_c(k,t)
\end{eqnarray}
\begin{eqnarray}\label{eq:acdot}
\dot{a}_c(k,t)&=&-iE_c(k)a_c(k,t)-E(t)\nabla_k a_c(k,t) 
+iE(t)\sum_j d^*_{jc}(k)a_j(t).
\end{eqnarray}
Here we assume only nearest neighbour hopping in the tight-binding approximation for the valence band (only ${a}_{j-1}$ and ${a}_{j+1}$ appear in the formula for $\dot{a}_{j}$). In solving Eq.~(\ref{eq:ajdot}) we take into consideration dynamics only due to the hopping in the lattice (the first two terms in Eq.~(\ref{eq:ajdot})) and the laser electric field (the third term).  Additionally, we neglect the valence electrons dynamics due to the valence-conduction dipole interaction (the last term in Eq.~(\ref{eq:ajdot})).

\begin{figure}[htbp]
\begin{center}
\includegraphics[width=0.6\linewidth]{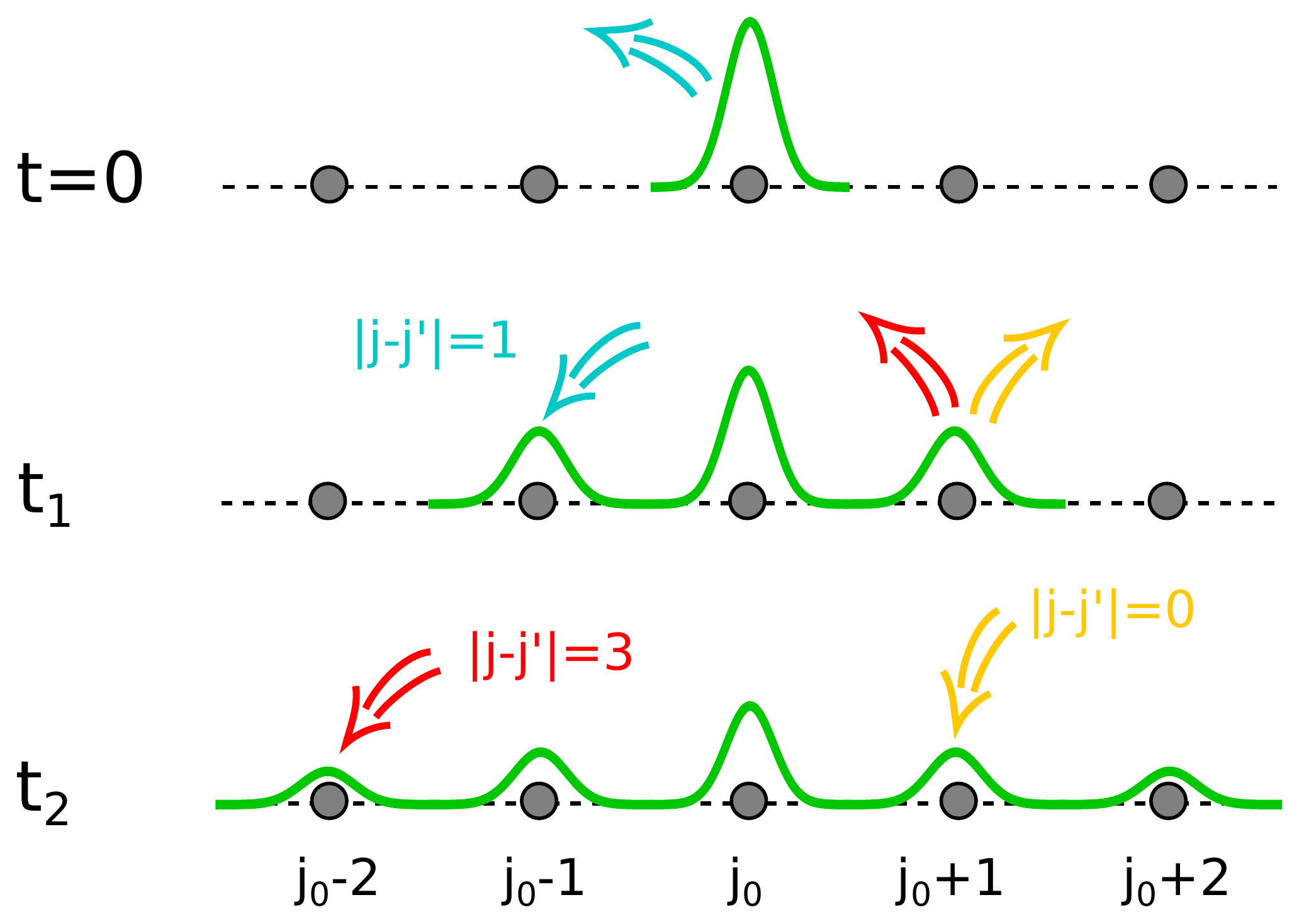}
\caption{Scheme of the electron excitation-recombination process in the periodic lattice. Electron's wave function, initially ($t=0$) centered at the $j_0$ ion, gradually spreads in the lattice. With colour arrows we show examples of electron excitation-recombination process with different relative displacements between the born atomic-parent site and the recombination atom site $|j-j'|$.}\label{fig:hhg_didactic}
\end{center}
\end{figure}

We note from Eq.~(\ref{eq:acdot}) that the coefficients $a_c(k,t)$ are directly related to $a_j(t)$, which, in turn, denote the localized place from which the electron will be excited from the valence to the conduction band.  This provides a localized picture quite distinct from the one obtained using the Bloch-Bloch approach \cite{Brabec}.
By neglecting the last term in Eq.~(\ref{eq:ajdot}) and solving it explicitly following~\cite{Dunlap}, we obtain
\begin{eqnarray}\label{eq:ajdyn}
a_j(t)&=&\sum_q a_q(0) e^{-iqa\eta(t)}(-\lambda)^{j-q} \nonumber \\ 
&\times&J_{q-j}(-2I_v[v^2(t)+u^2(t)]^{1/2})
\end{eqnarray}
where $q$ runs over all atomic sites, $a$ is lattice constant, $J_{q}$ are Bessel functions of $q$-th order and
\begin{equation}
\eta(t)=\int^t_0 dt' E(t'), 
\end{equation}
\begin{equation}
u(t)=\int^t_0 dt'\cos[a(\eta(t')-\eta(t))],
\end{equation}
\begin{equation}
v(t)=\int^t_0 dt'\sin[a(\eta(t')-\eta(t))],
\end{equation}
\begin{equation}
\lambda=\{[v(t)-iu(t)]/[v(t)+iu(t)]\}^{1/2}.
\end{equation}
We assume that the electron is initially localized at one atomic site $j_0$, i.e.~$a_q(0)=\delta_{qj_0}$. Later, due to the interatomic hopping and the acceleration due to the laser electric field, the electron's wave function spreads in the lattice following Eq.~(\ref{eq:ajdyn}). The width of the electron's wave function spread in the lattice at the end of the laser pulse depends on the hopping amplitude $I_v$, the lattice constant $a$, the laser electric field strength ${\mathcal E}_0$ and pulse duration.  We allow all coefficients $a_j(t)$ to acquire non-zero values during the laser pulse.
Equation~(\ref{eq:acdot}) is solved by following~\cite{Lewenstein1994, Brabec} and thereby we write  
\begin{eqnarray}
a_c(p,t)&=&i\sum_{j'}\int^t_0 dt'E(t')a_{j'}(t') d^*_{jc}(p+A(t')) \nonumber \\
&\times& e^{-i\int^t_{t'}E_c(p+A(t''))dt''}
\end{eqnarray}
here $A(t)$ is a laser vector potential $E(t)=-\frac{\partial A(t)}{\partial t}$, and $p=k-A(t)$ is the canonical momentum defined in the Brillouin zone shifted by $A(t)$, i.e. $\tilde{BZ}=BZ-A(t)$. 
Finally the time-dependent dipole moment $d(t)$ takes the form
\begin{eqnarray}\label{eq:dt}
d(t)&=&i|\tilde{w}_v|^2\sum_j\sum_{j'}\int^t_0 dt'\int_{\tilde{BZ}}dp\, a^*_{j}(t) d_{vc}(p+A(t))e^{i(p+A(t))\,x_j} \nonumber \\
&\times&e^{-i\varphi(p,t,t')} a_{j'}(t') d^*_{vc}(p+A(t'))e^{-i(p+A(t'))\,x_{j'}} E(t')+c.c. 
\end{eqnarray}
Where, $\varphi(p,t,t')=\int^t_{t'}E_c(p+A(t''))dt''$ is the accumulated phase of the electron in the conduction band. 
Equation~(\ref{eq:dt}) describes the harmonic emission originating from a single electron in a lattice.  Its interpretation is similar to harmonic emission in atoms.  In particular, the dipole radiation contains all the ``relative trajectory contributions'' of the electron wave function to the emitted harmonics from a solid. 
As depicted in Fig.~\ref{fig:hhg_didactic}, first, the electron located at the $j'$-th atomic site is excited from the valence to the conduction band. Second, it is accelerated within the conduction band by the laser field.  Finally, the electron has some probability of recombining with a different atom, located at $j$-th site (see the arrow pointing down in Fig.~\ref{fig:hhg_didactic}).  As a result, excess electron energy is emitted in the form of a high harmonic of the driving laser frequency.  Here it is assumed that the electron is initially localized at $j_0$ atomic site, as was mentioned above. 

To account for contribution of all electrons in a lattice, we multiply the dipole moment given by Eq.~(\ref{eq:dt}) by the total number of electrons, $N_e$. This is explained by cancellation of the two phases arising from shifting the initial Wannier function to another atomic site.  In particular,
let us consider another electron localized at a site $j_p=j_0+n$ at time $t=0$. In the dipole, Eq.~(\ref{eq:dt}), for any given $t$, $t'$ two additional phases will appear in comparison to the calculations where an electron starting at $j_0$ site is used: (i) $\exp[-i(A(t')-A(t))na]$ from the shift of $x_j$, $x_{j'}$ in (\ref{eq:dt}), and (ii) $\exp[ina\eta(t)]\exp[-ina\eta(t')]=\exp[-ina(A(t)-A(t'))]$ from the shift $j\rightarrow j+n$, $q\rightarrow q+n$ in $a^*_j(t)a_j'(t')$. Since these two phases cancel each other, the contribution from each electron in the system to the harmonic emission will be exactly the same.

\section{Results and discussion}

In this Section, we use the model developed in Section II to investigate harmonic emission of ZnO. The approach presented in this paper treats the HHG process within an atomistic-like approximation, assuming initial localization of an electron at one atomic site and gradual spread of its wave function in the lattice due to interatomic hopping and the influence of laser electric field. 
We consider two different directions for the laser field polarization $\Gamma-A$ and $\Gamma-M$~\cite{Kittel}.
In case of a narrow valence band (small values of $I_v$) the dynamics of the valence electrons is slow and the electron wave function does not spread much when the laser pulse is turned on. This means that only a few of the $a_j$ coefficients in the sum of Eq.~(\ref{eq:dt}) will have non-zero values.
For calculations in Sec. III.B, where we assume laser polarisation in the $\Gamma-A$ direction, we consider 15 atomic sites (ions) on both sides of $j_0$ (with this number we already obtain good convergence).  In the case of a wide valence band (Section III.A), where the laser polarisation is in the $\Gamma-M$ direction, the dynamics are much faster and we need to consider up to 1000 atomic sites to obtain convergence.

\subsection{Comparison of Bloch-Bloch and Wannier-Bloch models}

We begin by comparing the emitted HHG spectrum from the Wannier-Bloch approach, Eq.~(\ref{eq:dt}), with the spectrum obtained from the delocalized Bloch-Bloch approach. The Bloch-Bloch model was implemented following~\cite{Brabec}, with the cosine band structure approximated by a Taylor expansion up to the fourth order and integration in momentum space replaced by a saddle-point approximation.  Integration over the ionization time was done numerically using a Gaussian quadrature routine and the Fourier integral was performed as an FFT.

\begin{figure}[htbp]
\,\,\,\,\includegraphics[width=0.65\linewidth]{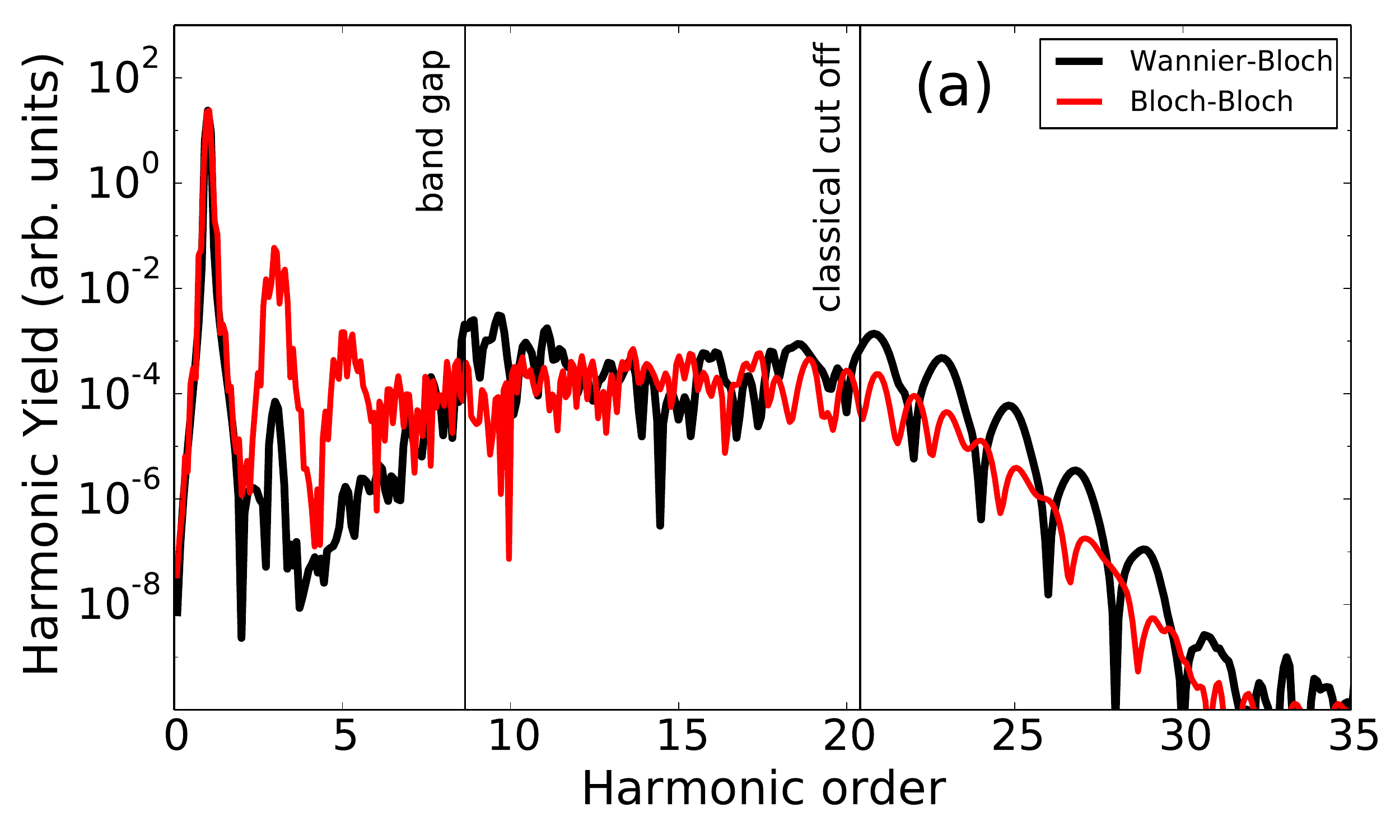}
\includegraphics[width=0.65\linewidth]{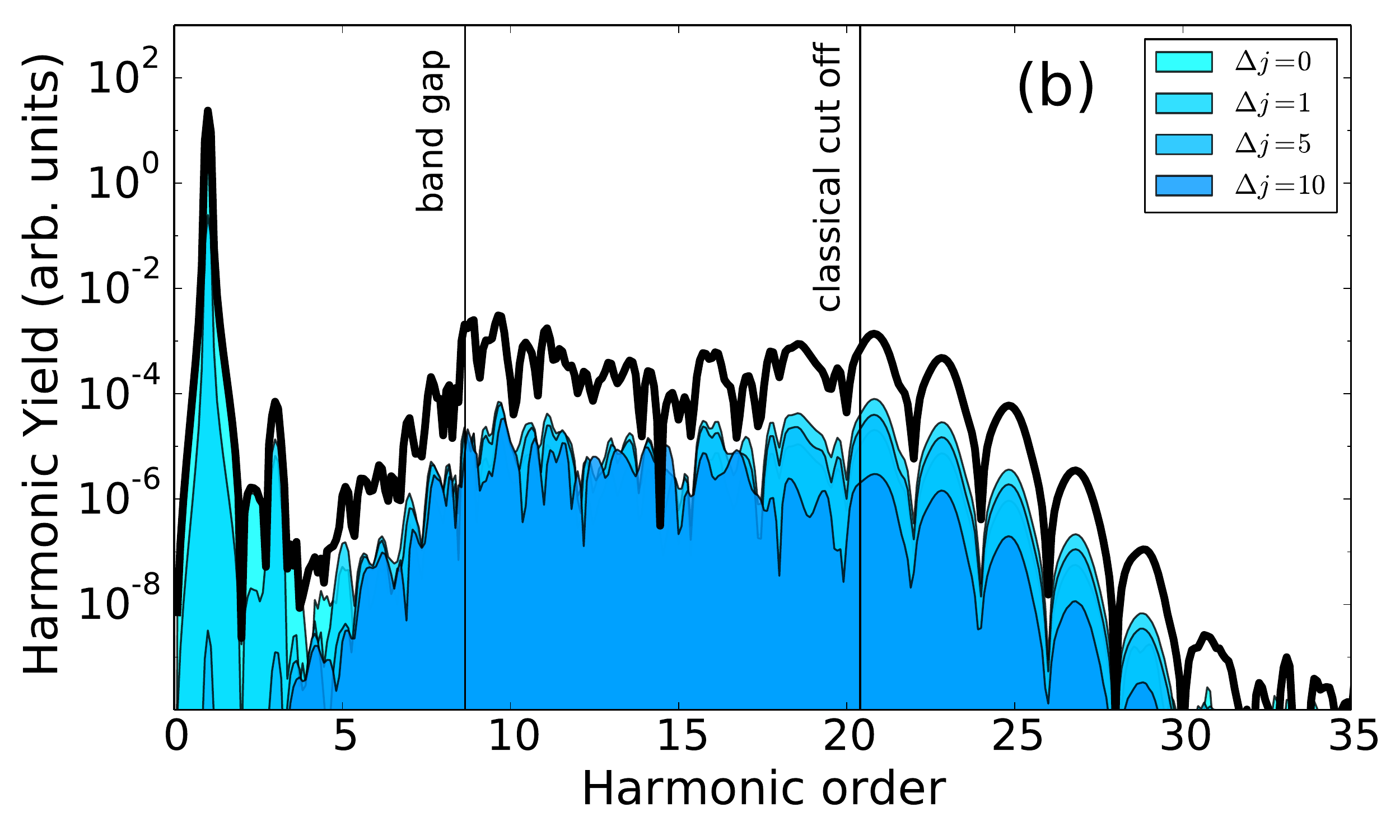}
\caption{Harmonic spectra comparison of the Wannier-Bloch approach (black) with the Bloch-Bloch approach (red) is shown in panel (a).  Decomposition of the harmonic contribution into different lengths of electron-recombination atomic-sites $\Delta j$ using Wannier-Bloch method is depicted in panel (b) for the corresponding color area plots. The calculations are carried out for laser polarization in $\Gamma-M$ direction of ZnO crystal. The laser parameters are: the carrier wavelength of $\lambda_0=3.25$ $\mu$m, laser intensity $I_0=3.15\times10^{11}$~W/cm$^{2}$, total number of laser cycles $N=10$ periods, and $\varphi_{\rm CEP}=0$. 
 }\label{fig:hhg_brabec}
\end{figure}

Figure~\ref{fig:hhg_brabec}(a) depicts the HHG spectrum computed using the same laser parameters as in Ref.~\cite{Vampa_bands}, i.e.~laser peak intensity $I_0=3.15\times10^{11}$~W/cm$^2$, carrier wavelength $\lambda_0=3.25~\mu m$ and pulse length of 10 laser periods (FWHM$\sim 53$ fs). For computational convenience we use a sine-squared envelope laser pulse (defined in Section II) instead of the Gaussian one used in \cite{Vampa_bands}. A simplified cosine-like band-structure in the $\Gamma-M$ direction, with the approximate parameters of Ref.~\cite{Vampa_bands} is used, i.e.~$E_g=0.1213$~a.u., $I_c=0.0449$~a.u., $I_v=-0.0464$~a.u. and lattice constant $a=5.32$~a.u.  Following 
~\cite{Brabec}, the dipole moment is assumed to be constant:  $d_{vc}(k)=3.46$~a.u. For details about the 1D-Bloch-Bloch calculation of Vampa et al. we refer to the Supplemental Material of Ref.~\cite{Brabec} and the subsequent article Ref.~\cite{Vampa_bands}.\\
Our approach exhibits good agreement with the Bloch-Bloch model, reproducing the plateau, the cutoff and the standard odd harmonic structure. The two spectra differ mainly in the low-order harmonics region, suggesting that localization (or recombination with the parent atom) may play greater importance in the production of low-order harmonics, as is confirmed in Fig. ~\ref{fig:hhg_brabec}(b).  Overall, this comparison confirms that the Wannier-Bloch picture reproduces the essential features of the  Bloch-Bloch model for the emitted harmonics.  \\
Figure ~\ref{fig:hhg_brabec}(b) shows the contributions to the harmonic spectra obtained from different components of the $j$, $j'$ sums in Eq.~(\ref{eq:dt}). The black line indicates the whole harmonic spectrum and the blue regions show individual contributions corresponding to
different distances,  $\Delta j=|j-j'|$, between the electron excitation (at $j'$) and recombination (at $j$) atomic sites. 
To calculate the contribution of a given $\Delta j$, we apply FFT on the part of the dipole $d(t)$ composed only of terms in Eq.~(\ref{eq:dt}) for which $|j-j'|$ is equal $\Delta j$.  
The relative length of displacement between the ionization and recombination sites is given by $\mathcal{D}_s=\Delta j\, a$, which we call the 
{\em length of Electron-recombination at a Relative Atom-center} (ERA). 

As  Fig.~\ref{fig:hhg_brabec}(b) shows, in certain parts of the spectrum, even $\Delta j=10$ paths contribute considerably to the total harmonic emission.  This is in clear contract to HHG in atomic gas, where the electron has to recombine with its parent atom, corresponding to $\Delta j=0$ contributions only.  In the next section, we attempt to understand why even relatively distant atomic sites can contribute significantly to the total emission spectrum in solids.

\subsection{Wannier-Bloch picture}
In order to further investigate the contribution of different ERA to the HHG spectrum, we calculate harmonic emission for a set of different laser-parameters.  This also allows us to establish under what conditions the Wannier-Bloch approach may be a more adequate description relative to the Bloch-Bloch one ~\cite{Brabec}.
Due to computational constraints in the HHG spectra calculations using the band-structure of Sec.~IIIA, here we focus on the narrower valence band case. For the latter we are able to scan a wider range of parameters and analyze the band structure influence on the different ERA contributions to the HHG. \\
In order to compute the HHG spectra, we fix the 
optical axis of ZnO (with polarization of the laser in the $\Gamma - A$ direction)~\cite{Brabec} and use:
$a=9.83$ a.u., $I_c=0.02175$ a.u., $I_v=-0.00295$ a.u.~and $E_g=0.1213$ a.u.  Also, following the formulation in~\cite{Brabec}, $d_{vc}(k)=\sqrt{\frac{E_p}{2 (E_c(k)-E_v(k))^2}}$,
with the Kane parameter $E_p$ set to 0.479.
\begin{figure}[h!]
\vspace{0.2cm}
\includegraphics[width=0.55\linewidth]{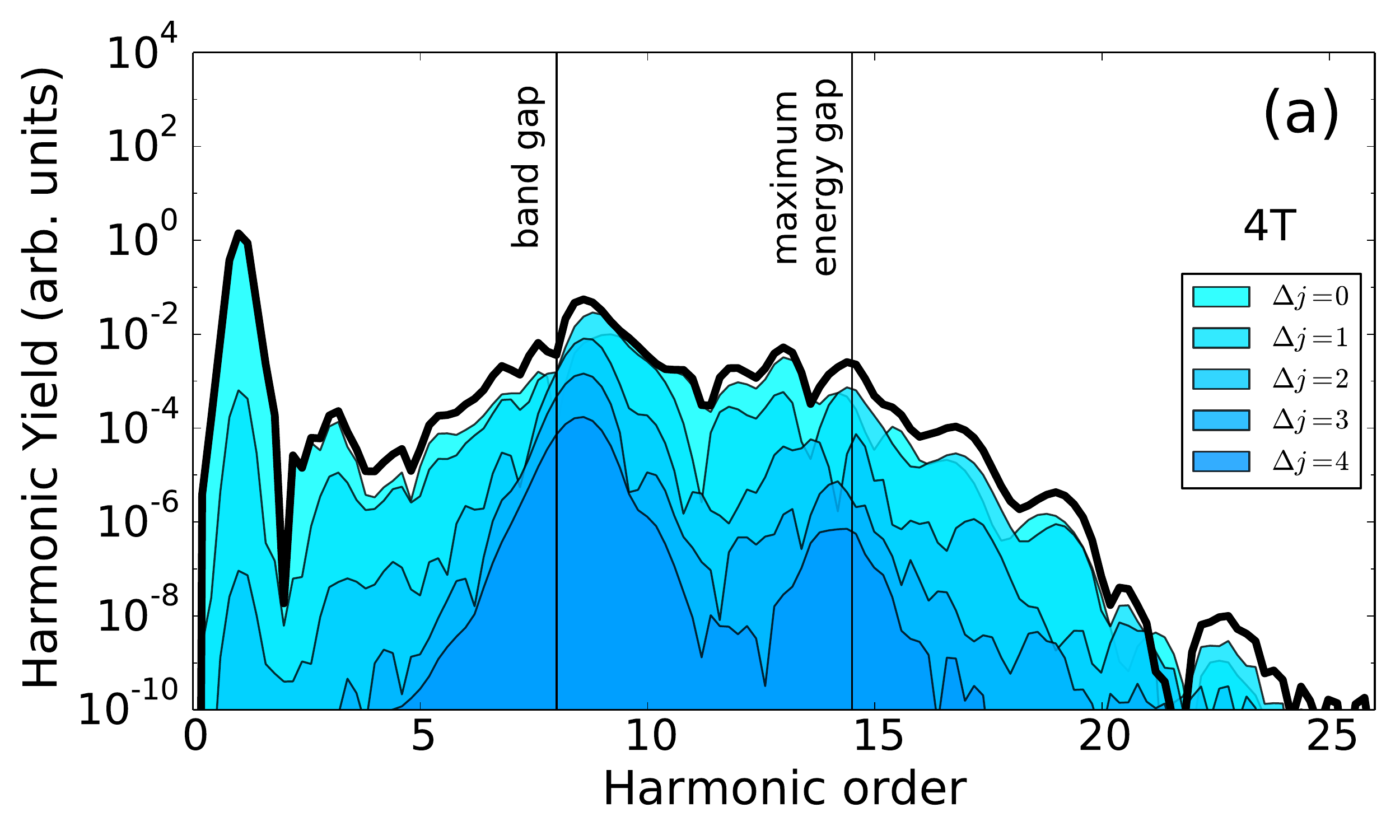}
\includegraphics[width=0.55\linewidth]{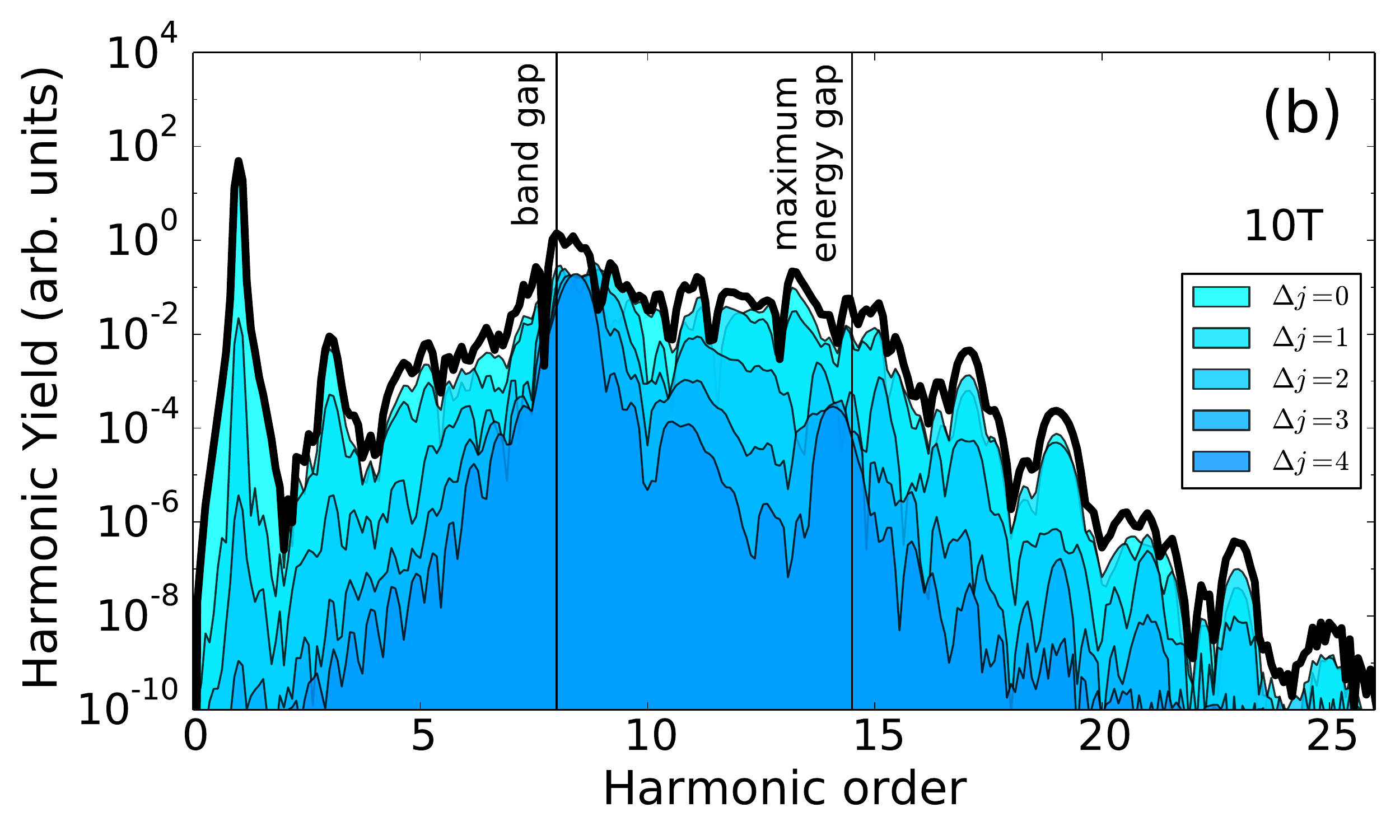}
\caption{Panels (a) and (b) depict the full harmonic spectra for two-different laser time-durations, $N=4$ and $10$ respectively (thick black lines), calculated within the Wannier-Bloch approach. Further, 
color regions show the harmonic contributions of the different relative electron-recombination atomic-sites in the spatially periodic lattice structure $\Delta j=0,\,1,...,\,4$. The black vertical lines (read from left to right) point out the band-gap harmonic and the expected cut-off harmonic orders, respectively. The laser pulse peak intensity is $I_0=5\times10^{11}$~W/cm$^2$, the carrier wavelength $\lambda_0=3.0~\mu m$ and its CEP is $\varphi_{\rm CEP}=\,0$ rad.}\label{fig:hhg_j}
\end{figure}
Figures~\ref{fig:hhg_j}(a) and~\ref{fig:hhg_j}(b) show the results of the emitted harmonic spectra
for two-different laser pulse durations, namely, $\Delta t_{b}=N\,T_0$, where $N=4$ and $N=10$ are the number of cycles, respectively, and $T_0$ the period of the laser field. 
It is observed that the highest contribution to the HHG spectrum comes from $\Delta j=0$, i.e.~for the case when the electron recombines to the same atomic-parent  site from where it was previously excited to the conduction band.
The longer the electron recombination displacement ${\mathcal D}_s$ is, the lower is its contribution to the HHG spectra. 
Both panels of Fig.~\ref{fig:hhg_j} show that, while the harmonics of odd orders decay very fast with $\Delta j$, there is 
a relatively large signal between the 8-th and 15-th harmonic which is preserved also for larger values of $\Delta j$. This signal corresponds to the energy gap between valence and conduction bands, which spreads
from about $\sim8\omega_0$ (band gap for $k=0$) to $14.5\omega_0$ (maximum energy gap for $k=\frac{\pi}{a}$). The greater contribution from large $\Delta j$ processes near the maximum
and minimum energy gaps can be understood as resulting from the high
density of possible interband transitions involving opposite band edges. 
Those transitions occur between states with narrowly defined momenta (near
the band extrema) which require broad spatial coherence as reflected in
large recombination lengths.\\
Figure ~\ref{fig:hhg_j} shows the typical features of HHG spectra:  namely odd harmonics are present, the signal is strongest for the  low-order harmonics (1st, 3rd), and a plateau region and a cutoff can be easily distinguished.  As would be expected for interband emission, the cutoff 
is located near the harmonic equivalent of the maximum energy difference between the conduction and valence bands. The region of the plateau exhibits pronounced interference structures - there is 
no clear recognition of even-odd harmonic symmetry. This behavior is also typical of the harmonic spectra in atoms in the limit of short pulses~\cite{Schmidt2012,Hentschel2001}. \\
The spectra in Fig~\ref{fig:hhg_j} can be compared with the experimental results shown in Fig.~3 of Ref.~\cite{Ghimire}, specifically for the crystal angle $0^{\circ}$. In this experimental data, 
up to the 13-th harmonic is distinguishable, which is close to the cut-off value obtained in our calculations.
As predicted by our model, a signal near the bandgap energy is observed in Ref.~\cite{Ghimire} (where, however, it is attributed to fluorescence effects).  In contrast to our results, in the experiment both odd and even harmonics appear in the spectrum. This is likely an effect
of a symmetry breaking in the 3D-ZnO lattice, which we do not take into account here.  

\begin{figure*}[h!]
\includegraphics[scale=0.42]{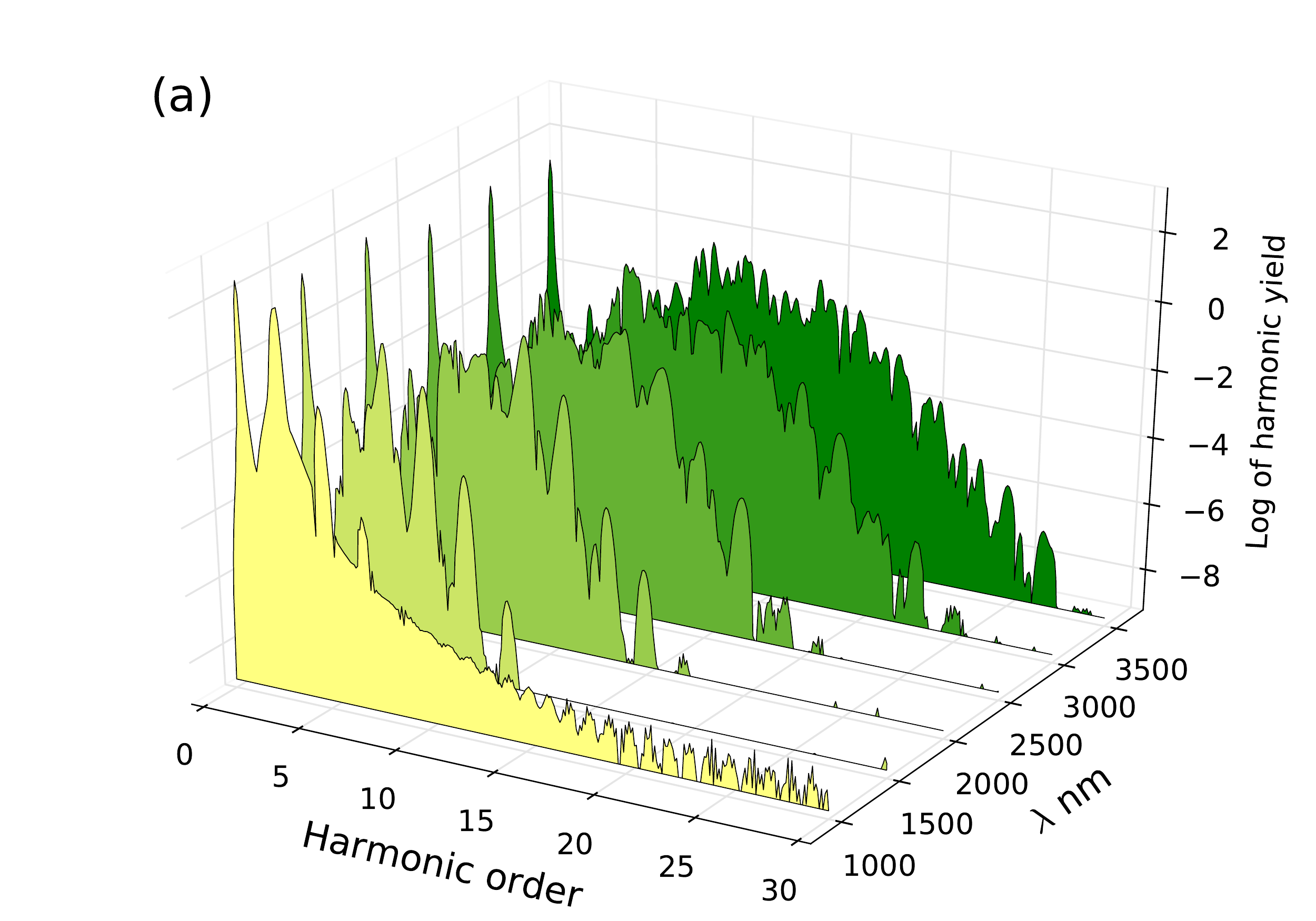}
\includegraphics[scale=0.40]{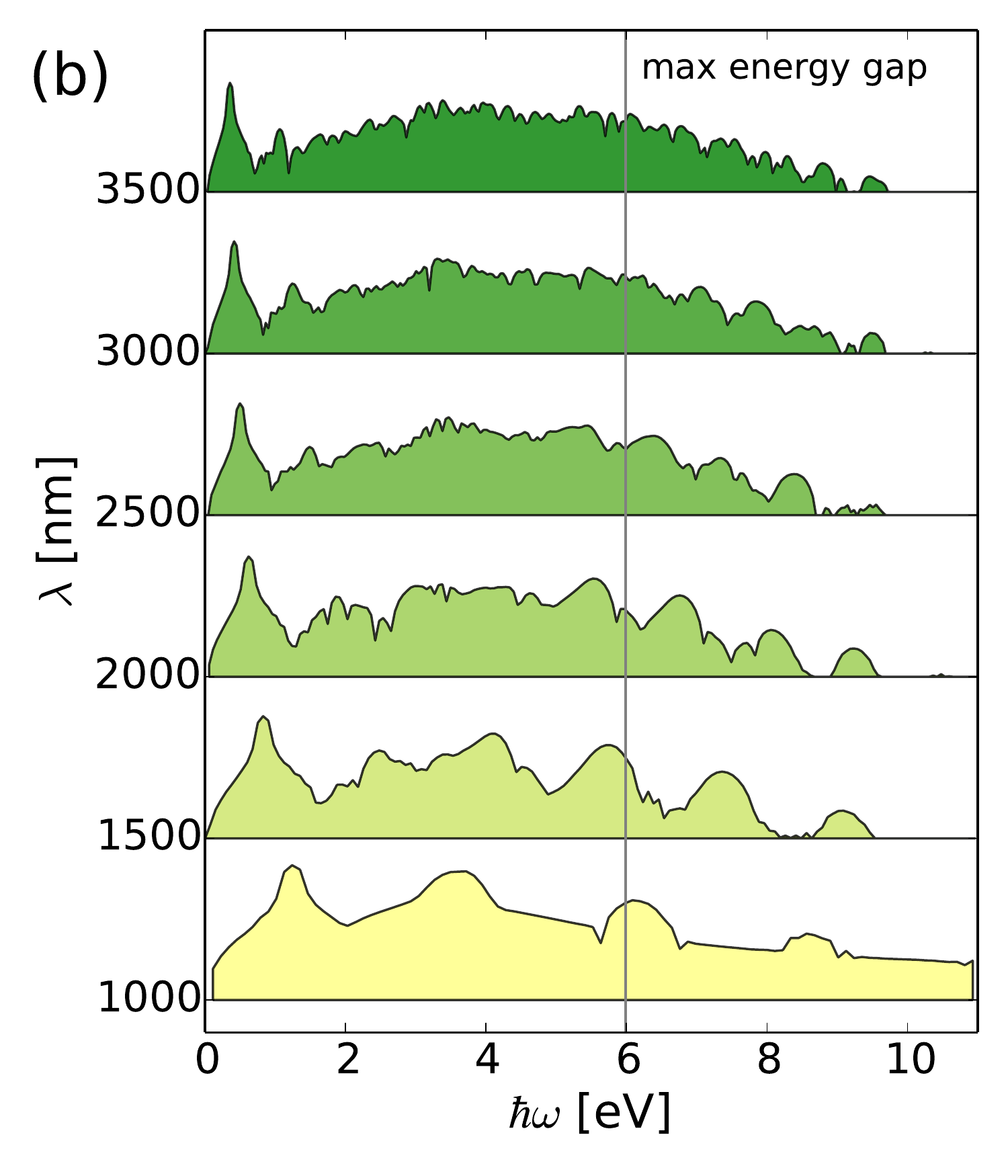}
\caption{(a) Harmonic spectra for different values of the carrier-laser wavelength, $\lambda_0$.
(b) Same as (a), but with energy units instead of harmonic order on the frequency axis. The grey line shows maximum emitted photon energy, which corresponds exactly to the maximum energy difference between the conduction and valence bands, ~$\Delta E_{vc} =6~e$V. Laser intensity was set to $I_0=5\times10^{11}$ W/cm$^{2}$, laser pulse length to 10 laser periods and $\varphi=0$ rad. }\label{fig:hhg_lambda}
\vspace{0.2cm}
\end{figure*}
To investigate how harmonic emission scales as a function of wavelength, we calculate the harmonic spectrum for different values of the laser wavelength, $\lambda_0$.  Note that the saddle point method was not used to solve the momentum integral in Eq.~(\ref{eq:dt}). The full integration is a reasonable choice while the energy variation of the dipole-radiation phase is comparable to the driving field frequency~\cite{DiChiara2009}. The argument is based on the fact that the dipole-radiation phase takes into account electron propagation in the conduction band. Nevertheless, for larger energy band-structure, the saddle point method is suitable for calculating the momentum integral in Eq.~(\ref{eq:dt}).  \\
The results are shown in Fig.~\ref{fig:hhg_lambda}. Plot (a) shows HHG spectra for
wavelengths in a range of 800-3500 nm. The frequency axis for each wavelength is scaled in $\omega_0=2\pi c/\lambda_0$ units. It is observed that the cut-off moves to lower harmonics while the wavelength decreases.
However, one may expect that the cut-off stays constant in terms of photon energy because of a well-defined {\em maximum energy difference} between the conduction and valence bands of about 6 eV.
This effect is shown on Fig. \ref{fig:hhg_lambda} (b), where the spectra of Fig. \ref{fig:hhg_lambda} (a) are replotted as a function of photon energy.  Hence, the cut-off of the harmonic spectrum is in good agreement with the value calculated from the band structure. From the so-called ``action phase" in Eq.~(\ref{eq:dt}), i.e. $\varphi(p,t,t')$, we can infer that the maximum harmonic energy produced in a solid lattice should be limited by the band dispersion relation (this result is consistent with prior findings~\cite{Brabec}).

\begin{figure}[h!]
\includegraphics[width=0.75\linewidth]{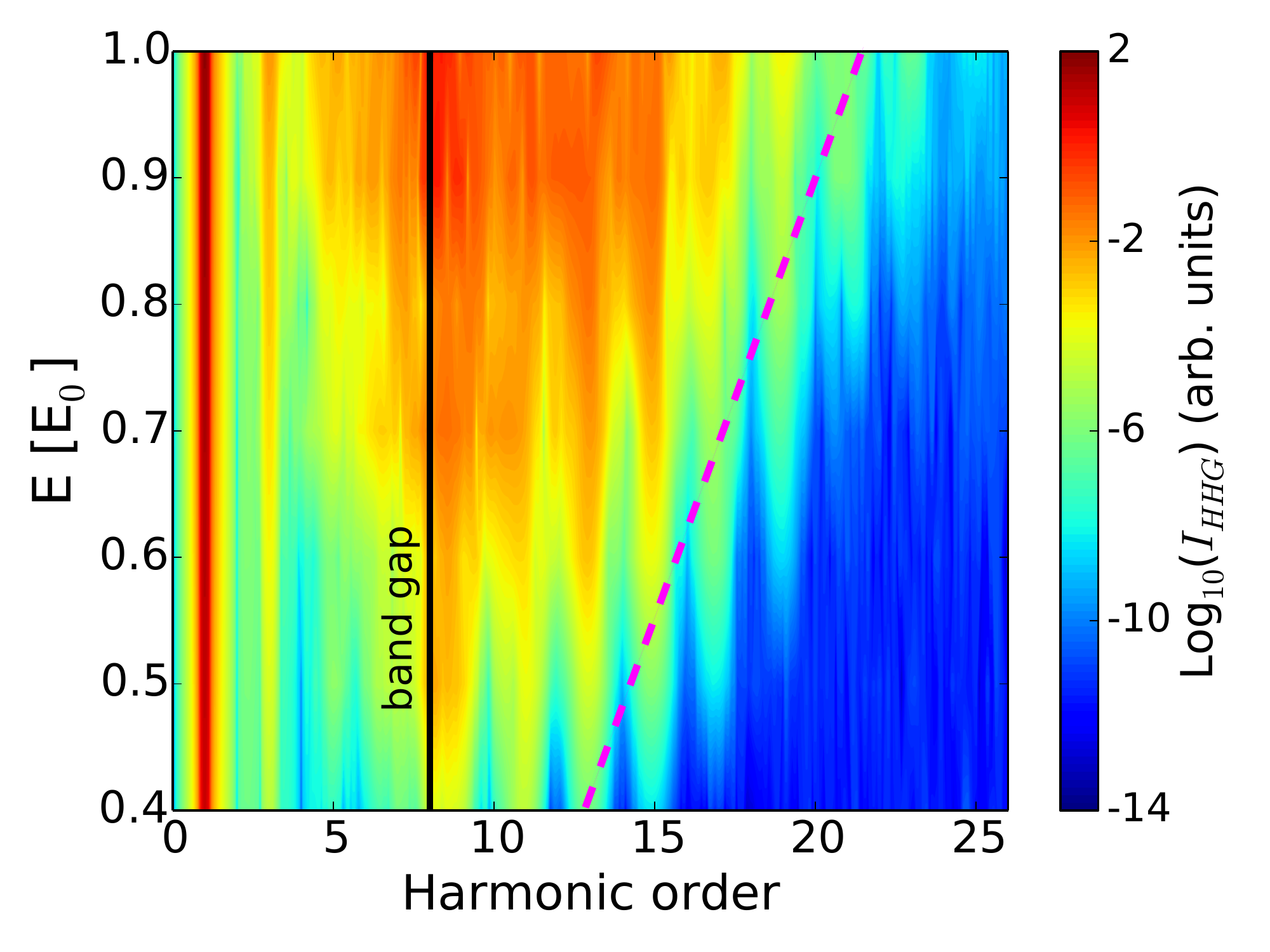}
\caption{Harmonic spectra as a function of electric field strength. The electric field axis is in units of $\mathcal{E}_0= 0.0038$~a.u., corresponding to the laser intensity $I_0=5\times10^{11}$ W/cm$^{2}$. The dashed magenta straight line denotes the estimated cut-off of the harmonic spectrum as a function of the electric field strength. The carrier-laser wavelength was set to $\lambda_0=3.0~\mu$m, 
laser pulse length to $N=10$ laser periods $T_0$, and $\varphi_{\rm CEP}=0$~rad. }\label{fig:hhg_intens}
\end{figure}
\begin{figure}[h!]
\includegraphics[width=0.65\linewidth]{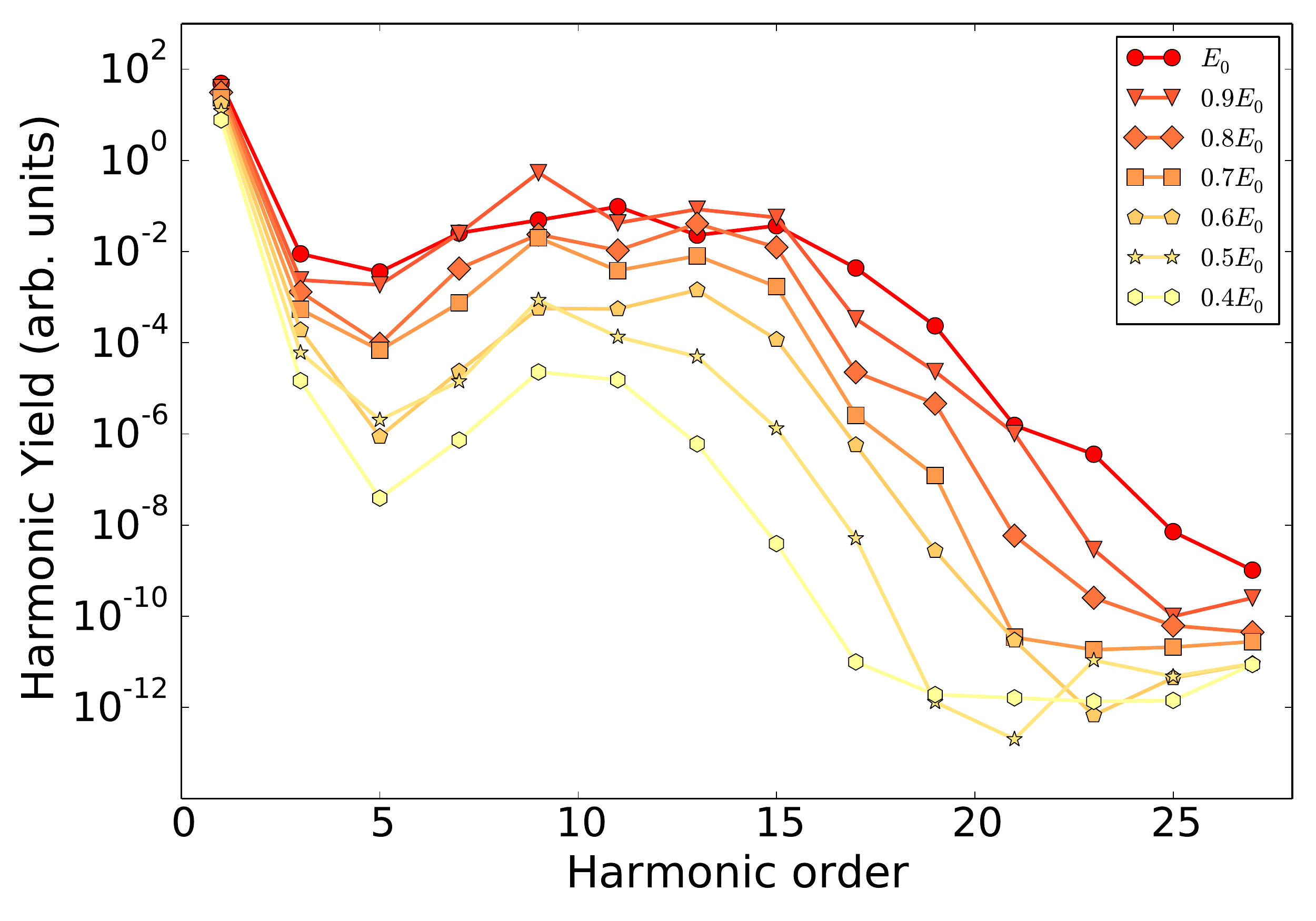}
\caption{Extracted peak spectral intensity of the odd harmonics from the full HHG spectra of Fig.~\ref{fig:hhg_intens}. Each line correspond to different strength of the electric field. Laser wavelength was set to $\lambda_0=3.0~\mu$m, 
laser pulse of the length of $N=10$ laser periods $T_0$. }\label{fig:hhg_intens2}
\end{figure}

Previously, it was found that HHG in solids scales linearly with the electric field strength ~\cite{DiChiara2009}.
Thereby, to further test the Wannier-Bloch model, we calculate HHG spectra for different values of the laser electric field.
Figure~\ref{fig:hhg_intens} shows HHG spectrum for electric field amplitudes, ${\mathcal E}_0$, in the range between 0.4-1.0${\mathcal E}_0$, where ${\mathcal E}_0=3.779\times10^{-3}$~a.u. corresponds to laser intensity $I_0=5\times10^{11}$ W/cm$^{2}$.  
As expected, decreasing the laser intensity results in shifting the cutoff to lower order harmonics. This can be seen more clearly in Fig.~\ref{fig:hhg_intens2} where the spectral intensity of the signal is plotted 
for all odd harmonics at each laser intensity. The fast decrease of the harmonic yield for
decreasing electric field strength at low ${\mathcal E}_0$ is a typical low-field
behavior in transitions induced by electric fields such as e.g. interband
Zener transitions. The decrease in cutoff with decreasing electric field strength was demonstrated experimentally in~\cite{DiChiara2009}.  Although, it is difficult to determine the exact position of the cutoff due to the few harmonics present for the lowest intensity cases, we trace in Fig.~\ref{fig:hhg_intens} an estimated straight line over the harmonic-electric field map. Excellent agreement of our cut-off model in terms of linear dependence on the electric field strength is observed as it is also experimentally demonstrated in Ref.~\cite{DiChiara2009}. 
For cases shown in Figs.~\ref{fig:hhg_lambda} and~\ref{fig:hhg_intens}, the relative contribution of longer displacements, $\mathcal{D}_s$, to harmonic emission is similar to what was observed in Fig.~\ref{fig:hhg_j}.  In particular, long relative displacements contribute significantly only to the harmonics close to the band gap energy.  Neither change of wavelength nor of laser intensity had a significant effect on the 
observed tendency. The situation is different for wider bands solids, as discussed in Sec. A.\\
For providing additional key information on localization of HHG in solids, our results agree in many aspects with prior experimental and theoretical observations ~\cite{Brabec,Vampa2015,Vampa_bands,Ghimire}.   In particular, ({\em i})  the HHG cutoff shows a dependence on the maximum energy difference between the valence and conduction bands as well as on the laser wavelength and peak intensity, ({\em ii}) for long laser pulses and few-cycle laser fields the model depicts the full odd and a continuum spectra, respectively, and ({\em iii}) we find a direct link between the emitted harmonic spectrum shape and the band-structure. 

\section{Concluding remarks}
 
By using localized atomic sites in the valence band and delocalized functions in the conduction band, our model has the closest parallels to harmonic generation from atomic gas.  As such, it allows one to access contributions of individual lattice sites, and hence assess the degree of localization of HHG in solids -- something that has previously been inaccessible.  
In particular, we can describe a process in which an electron initially localized at the $j'$-th atom in the valence band has a certain probability to be excited to the conduction band, where it is accelerated to a high energy before recombining either to the parent atom, at $j'$-th site, or (with different probability) to any other $j$-th atom in the lattice. \\
Different displacements of the electron recombination atomic-sites, i.e. $\Delta j=|j-j'|$, give different contributions to the harmonic spectrum. The approach developed here allows to extract all of these contributions. In particular, the main contribution was found to be given by $\Delta j=0$, or electron recombining at the same atomic site it was excited from. Especially for the case of narrow bands in the band structure, lower $\Delta j$
contribute by far the most to the harmonic spectrum, signifying substantial localization in the HHG process. On the other hand, we found enhanced contribution of high $\Delta j$ in case of wider valence and conduction bands.  This enhanced delocalization is likely due to increased mobility of the electrons in a lattice.  In all cases, distant neighbor contributions were highest near the band-gap energy.  This suggests that harmonic yield near the band gap energy should decline less (relative to other harmonics) with increasing ellipticity of laser light, since elliptical polarization tends to suppress local contributions.  \\
Note that our results by means of the Wannier-Bloch approach demonstrate a different interpretation than those presented by the Bloch-Bloch picture in ref.~\cite{Brabec}. While our model recreates the conventional atomic picture, the Bloch-Bloch model is based on the electron-hole pair recombination. Thereby, it is clear that both approaches, while predicting similar total HHG spectra, provide different interpretations of the HHG process. \\
Our results suggest that it should be possible to control the localization of the HHG process by varying experimental parameters.  Hence, by quantifying site specific contributions, our work paves the way to controlling HHG efficiency, creation of attosecond pulses and imaging of the electronic wave function in a crystal lattice \cite{Reis2016}.



\begin{acknowledgments}
E.N.O. acknowledges the support from the doctoral stipend ETIUDA of the National Science Centre
according to decision DEC-2015/16/T/ST3/00266. A.~C.~and M.~L.~acknowledge MINECO (Nacional Plan Grant FOQUS FIS2013-46768-P, Severo Ochoa Grant/SEV-2015-0522), AGAUR  grant 2014 SGR
874,  Fundaci\'o Cellex Barcelona, and ERC AdG OSYRIS.
 We also acknowledge the support from the European Union's Horizon 2020 research and innovation programme under grant
agreement no 654148 Laserlab-Europe and the Ministerio de Econom\'ia y Competitividad of 
 Spain (FURIAM project FIS2013-47741-R and DYNAMOLS FIS2013-41716-P).  
 This work was supported by the project ELI--Extreme Light Infrastructure--phase 2
(CZ.02.1.01/0.0/0.0/15\_008/0000162 ). A.L. acknowledges
partial support of Max Planck Postech/Korea Research Initiative Program
[Grant No 2011-0031558] and the Max Planck Center for Attosecond Science
(MPC-AS).
\end{acknowledgments}

\bibliography{BiblioE}

\end{document}